# Substitution Effects on FeSe Superconductor

Yoshikazu Mizuguchi[1,2,3], Fumiaki Tomioka[1,3], Shunsuke Tsuda[1,3] , Takahide Yamaguchi[1,3] and Yoshihiko Takano[1,2,3]

[1]*National Institute for Materials Science, 1-2-1, Sengen, Tsukuba, 305-0047, Japan*

[2]*University of Tsukuba, 1-1-1, Tennodai, Tsukuba, 305-8577, Japan*

[3]*JST, TRIP, 1-2-1, Sengen, Tsukuba, 305-0047, Japan*

We have investigated the effect of atomic substitutions in the FeSe system, which exhibits the simplest crystal structure among the iron-based superconductors. An enhancement of the superconducting transition temperature $T_c$ was observed with the substitution of S or Te for Se; the $T_c$ increased with S substitution by up to 20 %, and also increased with Te substitution up to 75 %. In contrast, Co or Ni substitutions for the Fe site significantly suppressed superconductivity. In this work we present a detailed description of the substitution technique employed to determine $T_c$ in the FeSe system.

## 1. Introduction

The discovery of superconductivity at a high temperature of 26 K (raised to 43 K under high pressure) in ZrCuSiAs-type structured $LaFeAsO_{1-x}F_x$ was received with considerable surprise.[1,2] To date, various iron-based superconductors with FeAs layers, analogous to LaFeAsO, have been reported. The substitution of small ions for La raised the $T_c$ above 50 K.[3 ~ 6] Kito *et al.* synthesized a superconducting phase of F-free $NdFeAsO_{1-x}$ using a high-pressure synthesis technieque.[7] Doping Co into the Fe site of the parent compound also induces superconductivity in the LaFeAsO system.[8] ZrCuSiAs-type structured SrFeAsF also shows a superconducting transition with Co doping.[9] Superconductivity at 38 K in $Ba_{1-x}K_xFe_2As_2$, which has a $ThCr_2Si_2$-type structure, was reported.[10] Co or Ni doping to the parent compound induces superconductivity as in the case of ZrCuSiAs-type structured system.[11,12] The $T_c$ is also lower than that of $Ba_{1-x}K_xFe_2As_2$. Another related compound is

$Li_{1-x}FeAs$. This compound also exhibits the superconducting transition with a $T_c \sim 18$ K.[13)]

Recently, superconductivity at 8 K in PbO-type FeSe was reported.[14)] FeSe has the simplest structure among the iron-based superconductors. This compound shows surprisingly a high $T_c^{onset}$ of 27 K under high pressure[15)]. X-ray and neutron diffraction patterns indicated a structural phase transition from tetragonal to orthorhombic.[16)] In $^{77}$Se NMR measurements, the nuclear-spin lattice relaxation ratio $1/T_1$ showed $T^3$ behavior below $T_c$ without a coherence peak.[17)] Similar behaviors have been reported in some of the iron-based superconductors. Te substitution for Se enhances $T_c$ in spite of a negative chemical pressure effect.[18,19)] Density functional calculations of FeS, FeSe and FeTe indicated that the strength of spin density wave (SDW) in FeTe and the possibility of a higher $T_c$ in doped FeTe or Fe(Se,Te) alloy compared to FeSe.[20)] The enhancement of $T_c$ in Te-substituted FeSe may be explained by the results of the density functional calculations.

The relationship between the external pressure effect and chemical pressure effect is unclear. To elucidate the mechanism for changes in $T_c$, we investigated the substitution of S (smaller than Se) and Te (lager than Se) for Se. We also investigated Co- and Ni-doping effects on the Fe site.

## 2. Experiments

Polycrystalline samples of $FeSe_{1-x}S_x$ ($x = 0 \sim 0.5$), $FeSe_{1-x}Te_x$ ($x = 0 \sim 1$), $Fe_{1-x}Co_xSe$ ($x$ = 0.05, 0.1 and 0.2) and $Fe_{1-x}Ni_xSe$ ($x$ = 0.05, 0.1 and 0.2) were prepared using a solid state reaction technique with heating temperatures of 680 ~ 800ºC. We characterized the obtained samples by powder X-ray diffraction using CuKα radiation. Resistivity measurements were performed using a four terminal method from 300 to 2 K. Temperature dependence of magnetization was measured using a SQUID magnetometer at $H$ = 10 Oe.

## 3. Results and discussion

### *3.1 X-ray diffraction*

Figure 1 shows the X-ray diffraction patterns of FeTe, $FeTe_{0.92}$, $FeSe_{0.5}Te_{0.5}$, FeSe, $FeSe_{0.8}S_{0.2}$, $Fe_{0.8}Co_{0.2}Se$ and $Fe_{0.8}Ni_{0.2}Se$. The peaks were well indexed using a space group of *P*4/*nmm*. The obtained lattice constants *a*, *c* and *V* are plotted in Fig. 2(a), (b) and (c), respectively. The values of lattice constants of all the samples are listed in Table I.

The expected X-ray diffraction pattern for $FeTe_{0.92}$ was obtained and no impurity phases were found. In FeTe, the peaks were indexed using *P*4/*nmm* except for an unidentified small peak at $2\theta \sim 32^{\circ}$. No hexagonal phase was detected for $FeTe_{0.92}$ and FeTe.

In $FeSe_{1-x}Te_x$, both *a* and *c* increased corresponding to the ionic radii of Te. $FeSe_{0.5}Te_{0.5}$ and $FeSe_{0.25}Te_{0.75}$ are single phases. For $FeSe_{0.75}Te_{0.25}$, the peaks of the tetragonal phase split. There are two phases both with the space group of *P*4/*nmm*. Similar results have been reported for $FeSe_{0.85}Te_{0.15}$ and $FeSe_{0.75}Te_{0.25}$.[18)] This split can be explained by a phase separation of the different Te concentrations.

In $FeSe_{1-x}S_x$, both *a* and *c* decreased corresponding to the ionic radii of Se. The change of *c* against *x* seems to saturate above $x = 0.3$; there may exist a solid solubility limit of S. There are peaks of the hexagonal phases ($P6_3/mmc$) (minor phases) indicated with asterisks in Fig. 1.

In $Fe_{1-x}Co_xSe$, both *a* and *c* decreased with increasing Co substitution. This is the same results observed in the other Co-doped iron-based superconductors.[8,21)] In contrast, *a* increased, while *c* decreased with increasing Ni substitution in $Fe_{1-x}Ni_xSe$.

### *3.2 Resistivity and magnetization*

#### *3.2.1 $FeSe_{1-x}S_x$*

Figure 3(a) shows the temperature dependence of resistivity for $FeSe_{1-x}S_x$ and Fig. 3 (b) is the data on a larger scale for low temperatures. The onset, mid-point and zero-point temperatures of the $T_c$ ($T_c^{onset}$, $T_c^{mid}$ and $T_c^{zero}$) are plotted in Fig. 3(c). The $T_c^{onset}$ indicated by arrows in Fig. 3(b) is defined as the temperature when the resistivity deviates from the linear temperature dependence. The $T_c^{mid}$ is defined as the temperature where the resistivity is half of the resistivity at $T_c^{onset}$. The $T_c^{onset}$ and the $T_c^{zero}$ increased for $x \leq 0.2$ and decreased for $x \geq 0.3$. The highest $T_c^{onset}$ was 15.5 K. An S ion has the same valence with a Se ion and has a smaller ionic radius than Se. Therefore, the S substitution corresponds to positive chemical pressure. The enhancement of $T_c$ in $FeSe_{1-x}S_x$ is consistent with the result of pressurized FeSe. However, the highest $T_c$ is much smaller than that of pressurized FeSe.

Figure 4 shows the temperature dependence of magnetization for $FeSe_{1-x}S_x$. The superconducting transition temperatures $T_c^{mag}$ were estimated to be 8.6 and 9.9 K for $x = 0.1$ and 0.2. The shielding volume fractions were calculated using the value of magnetization (ZFC) at 2 K and estimated to be 27 % and 68 % for $x = 0.1$ and 0.2. The volume fraction for $x = 0.2$ is about three times lager than that of FeSe. The S substitution seems to stabilize superconducting states in FeSe.

### 3.2.2 *FeSe*$_{1-x}$*Te*$_x$

Figure 5(a) is the temperature dependence of resistivity for $FeSe_{1-x}Te_x$ and Fig. 5 (b) is the data on a larger scale for low temperatures. The onset, mid-point and zero-point temperatures of the $T_c$ ($T_c^{onset}$, $T_c^{mid}$ and $T_c^{zero}$) are plotted in Fig. 5(c). The $T_c$ increased with Te substitution for $x \leq 0.75$, as reported in refs. 17 and 18. The highest $T_c^{onset}$ and $T_c^{zero}$ are 15.3 K for $x = 0.25$, and 11.8 K for $x = 0.75$. Superconducting transition widths for $x = 0.5$ and 0.75 are shaper than that for $x = 0.25$ or FeSe. For FeTe, superconductivity disappeared and a strong anomaly, which corresponds to a structural phase transition, appeared on resistivity around 70 K.

Figure 6 shows the temperature dependence of magnetization for $FeSe_{1-x}Te_x$. $T_c^{mag}$ of superconducting transition were estimated to be 13.5, 10.5 and 9.5 K for $x = 0.25$, 0.5 and 0.75, respectively. The shielding volume fractions were estimated to be about 40 % for $x = 0.25$ and to be almost 100 % for $x = 0.5$ and 0.75. The $T_c^{mag}$ is the highest for $x = 0.25$. However, the volume fraction is smaller than that for $x = 0.5$ or 0.75.

### 3.2.3 *Fe*$_{1-x}$*Co*$_x$*Se, Fe*$_{1-x}$*Ni*$_x$*Se*

Figure 7 shows the temperature dependence of resistivity for $Fe_{1-x}Co_xSe$. Superconducting transition was suppressed by the substitution of Co for Fe. The $T_c^{onset}$ were estimated to be 10 and 5 K for $x = 0.05$ and 0.1. The resistivity at room temperature decreases with increasing Co concentration. The gradient of the temperature dependence of resistivity decreases with increasing Co concentration. This behavior may be due to impurity scattering in the Fe plane.

Figure 8 shows the temperature dependence of resistivity for $Fe_{1-x}Ni_xSe$. Superconducting transition was suppressed by the substitution of Ni for Fe. The $T_c^{onset}$ were estimated to be 10 K for $x = 0.05$. The resistivity at room temperature increased with increasing Ni substitution.

### 3.3 *Discussion for the overall results*

The $T_c$ of FeSe was enhanced with substitution of both S and Te for Se. Both substitutions also enhance the superconducting volume fraction. Contrary to our expectation, the $T_c$ of pure FeSe is the lowest in this system and the transition is broader for pure FeSe. This result suggests that the substitution of S or Te for the Se site stabilizes superconductivity. Unlike other iron-based superconductors, superconducting FeSe shows a structural phase transition to orthorhombic. The coexistence of superconductivity and the structural phase transition may be the cause of the lower $T_c$, the broad transition and the smaller superconducting volume fraction in FeSe. The substitution of S or Te for Se may suppress the structural phase transition competing with superconductivity.

$T_c^{zero}$s of $FeSe_{1-x}S_x$, FeSe and $FeSe_{1-x}Te_x$ in Fig. 3(c) and 5(c) show that the increase of $T_c^{zero}$ in $FeSe_{1-x}Te_x$ is larger than that in $FeSe_{1-x}S_x$. This seems to indicate that the chemical pressure effect is not exactly consistent with the external pressure effect for FeSe. Another factor, other than the chemical pressure effect determines the $T_c$, in this system.

To clarify the factor, which determines $T_c$ in this system, we compared the Fe(Se,Te) system and the LaFe(As, P)O system. Figure 9 shows the temperature dependence of resistivity for FeTe, FeSe, LaFeAsO[1)] and LaFePO[22)]. Both FeSe and LaFePO show metallic behavior and superconductivity at low temperatures without any substitution. In contrast, FeTe and LaFeAsO, which show no superconducting transition, exhibit superconductivity by elementary substitutions, for example, $LaFeAsO_{1-x}F_x$, $FeTe_{1-x}Se_x$ and $FeTe_{1-x}S_x$[18, 19, 23)].

The $T_c$ of $LaFeAsO_{1-x}F_x$ is much higher than that of LaFePO. In this respect, FeTe is expected to become a superconductor at a higher $T_c$ compared to FeSe. This possibility was also mentioned in a previous theoretical study.[20)] A density functional calculation study indicated that the higher stability of the SDW state in FeTe compared to FeSe, suggesting that chemically doped FeTe could have a higher $T_c$ than that of FeSe. The prediction is consistent with our results. The enhancement of $T_c$ in $FeSe_{1-x}Te_x$ may be explained by the density functional calculation. However, the highest $T_c$ in $FeSe_{1-x}Te_x$ is smaller than that of pressurized FeSe. Other methods can induce superconductivity with a higher $T_c$ in FeTe.

Contrary to the substitution for the Se site, the substitution for the Fe site strongly suppressed superconductivity. The substitution for the Fe site, which is essential for the occurrence of superconductivity, may induce randomness, and suppress superconductivity.

## Acknowledgment

This work was partly supported by Grant-in-Aid for Scientific Research (KAKENHI).

**94** (2009) 012503.

Table I. The values of lattice constants. For $FeSe_{0.75}Te_{0.25}$, there are two phases with high (1) and low (2) concentration of Te.

| Sample | $a$ (Å) | $c$ (Å) | $V$ (Å$^3$) |
| --- | --- | --- | --- |
| FeTe | 3.8097(25) | 6.2756(17) | 91.08(13) |
| $FeTe_{0.92}$ | 3.822(2) | 6.292(3) | 91.9(1) |
| $FeSe_{0.25}Te_{0.75}$ | 3.814(1) | 6.157(2) | 89.56(7) |
| $FeSe_{0.5}Te_{0.5}$ | 3.7909(5) | 5.957(1) | 85.61(3) |
| $FeSe_{0.75}Te_{0.25}$ (1) | 3.7989(3) | 5.968(4) | 86.13(5) |
| $FeSe_{0.75}Te_{0.25}$ (2) | 3.7815(16) | 5.617(3) | 80.32(8) |
| FeSe | 3.7696(6) | 5.520(1) | 78.44(3) |
| $FeSe_{0.9}S_{0.1}$ | 3.763(2) | 5.503(7) | 77.9(1) |
| $FeSe_{0.8}S_{0.2}$ | 3.754(1) | 5.442(7) | 76.7(1) |
| $FeSe_{0.7}S_{0.3}$ | 3.750(2) | 5.399(8) | 75.9(1) |
| $FeSe_{0.6}S_{0.4}$ | 3.747(13) | 5.342(27) | 75.0(7) |
| $FeSe_{0.5}S_{0.5}$ | 3.732(2) | 5.381(5) | 75.0(1) |
| $Fe_{0.95}Co_{0.05}Se$ | 3.7675(10) | 5.515(2) | 78.27(6) |
| $Fe_{0.9}Co_{0.1}Se$ | 3.7637(4) | 5.5043(8) | 77.97(2) |
| $Fe_{0.8}Co_{0.2}Se$ | 3.7562(7) | 5.483(1) | 77.36(3) |
| $Fe_{0.95}Ni_{0.05}Se$ | 3.7714(6) | 5.508(1) | 78.34(3) |
| $Fe_{0.9}Ni_{0.1}Se$ | 3.7713(3) | 5.503(1) | 78.27(2) |
| $Fe_{0.8}Ni_{0.2}Se$ | 3.7750(9) | 5.489(5) | 78.22(8) |

**Figure captions**

Fig. 1. X-ray diffraction pattern of FeTe, $FeTe_{0.92}$, $FeSe_{0.5}Te_{0.5}$, FeSe, $FeSe_{0.8}S_{0.2}$, $Fe_{0.8}Co_{0.2}Se$ and $Fe_{0.8}Ni_{0.2}Se$. Asterisks indicate hexagonal phase and a triangle indicate impurity phase. Other peaks were determined as tetragonal phase.

Fig. 2. (Color online) Substitution dependence of lattice constants (a)*a*, (b)*c* and (c)*V*. Open diamonds, open squares, filled diamonds, and filled triangles correspond to $Fe_{1-x}Ni_xSe$, $Fe_{1-x}Co_xSe$, $FeSe_{1-x}Te_x$, and $FeSe_{1-x}Te_x$, respectively. The values are shown in table 1.

Fig. 3. (Color online) (a)Temperature dependence of resistivity for $FeSe_{1-x}S_x$. (b) Enlargement of low temperatures. Arrow indicates $T_c^{onset}$. (c) Substitution dependence of $T_c^{onset}$ (squares), $T_c^{mid}$ (circles) and $T_c^{zero}$ (diamonds) for $FeSe_{1-x}S_x$.

Fig. 4. (Color online) Temperature dependence of magnetization for $FeSe_{1-x}S_x$. Open and filled marks correspond to the data obtained by zero field and field cooling, respectively.

Fig. 5. (Color online) (a)Temperature dependence of resistivity for $FeSe_{1-x}Te_x$. (b) Enlargement of low temperatures. Arrow indicates $T_c^{onset}$. (c) Substitution dependence of $T_c^{onset}$ (squares), $T_c^{mid}$ (circles) and $T_c^{zero}$ (diamonds) for $FeSe_{1-x}Te_x$.

Fig. 6. (Color online) Temperature dependence of magnetization for $FeSe_{1-x}Te_x$. Open and filled marks correspond to the data obtained by zero field and field cooling, respectively.

Fig. 7. (Color online) Temperature dependence of resistivity for $Fe_{1-x}Co_xSe$.

Fig. 8. (Color online) Temperature dependence of resistivity for $Fe_{1-x}Ni_xSe$.

Fig. 9 (Color online) Temperature dependence of resistivity for FeTe (the actual Te concentration is ~0.92)[23)], FeSe (the actual Se concentration is ~0.92)[15)], LaFeAsO[1)] and LaFePO[22)].

Fig. 1

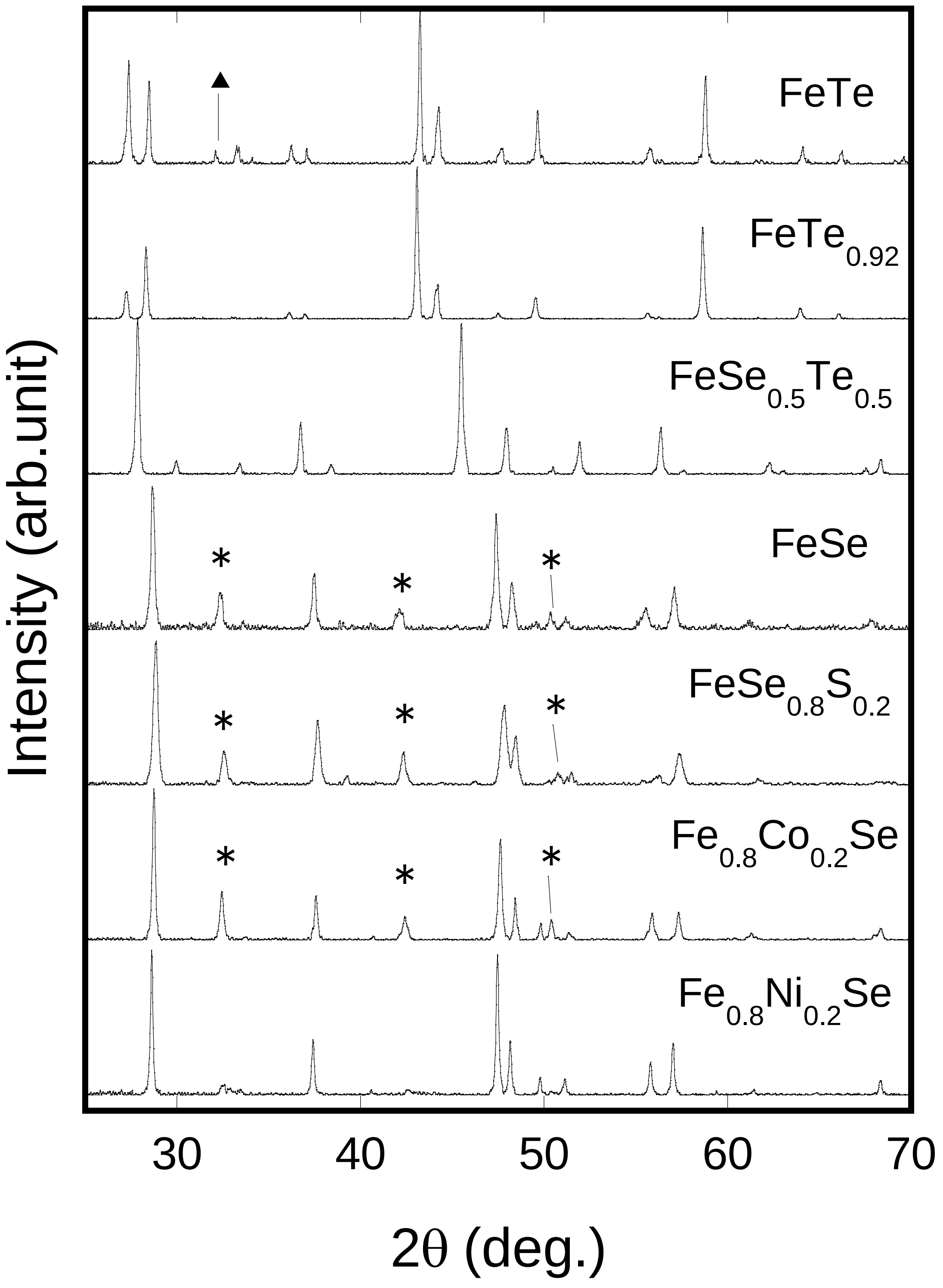

Fig. 2

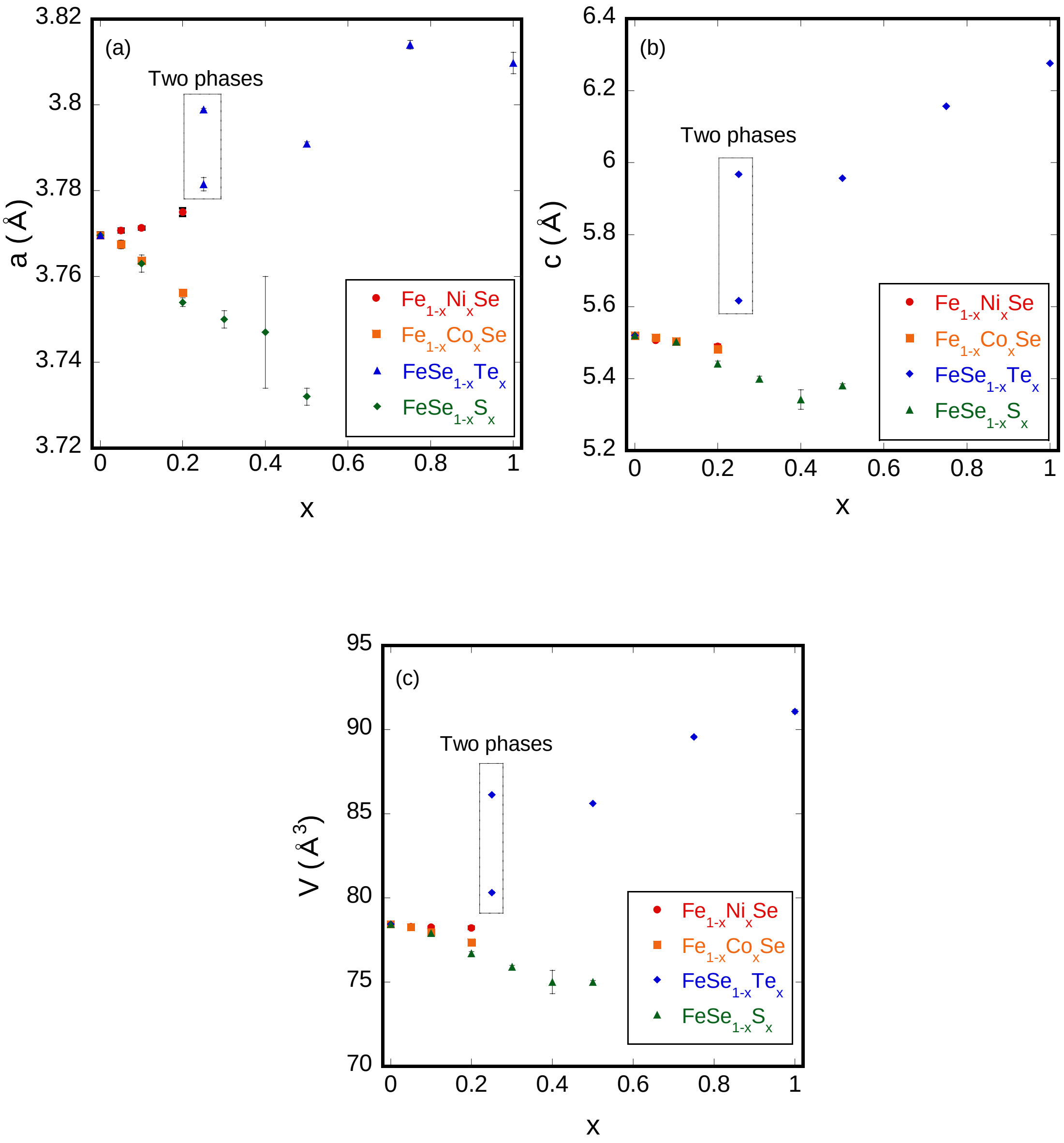

Fig. 3

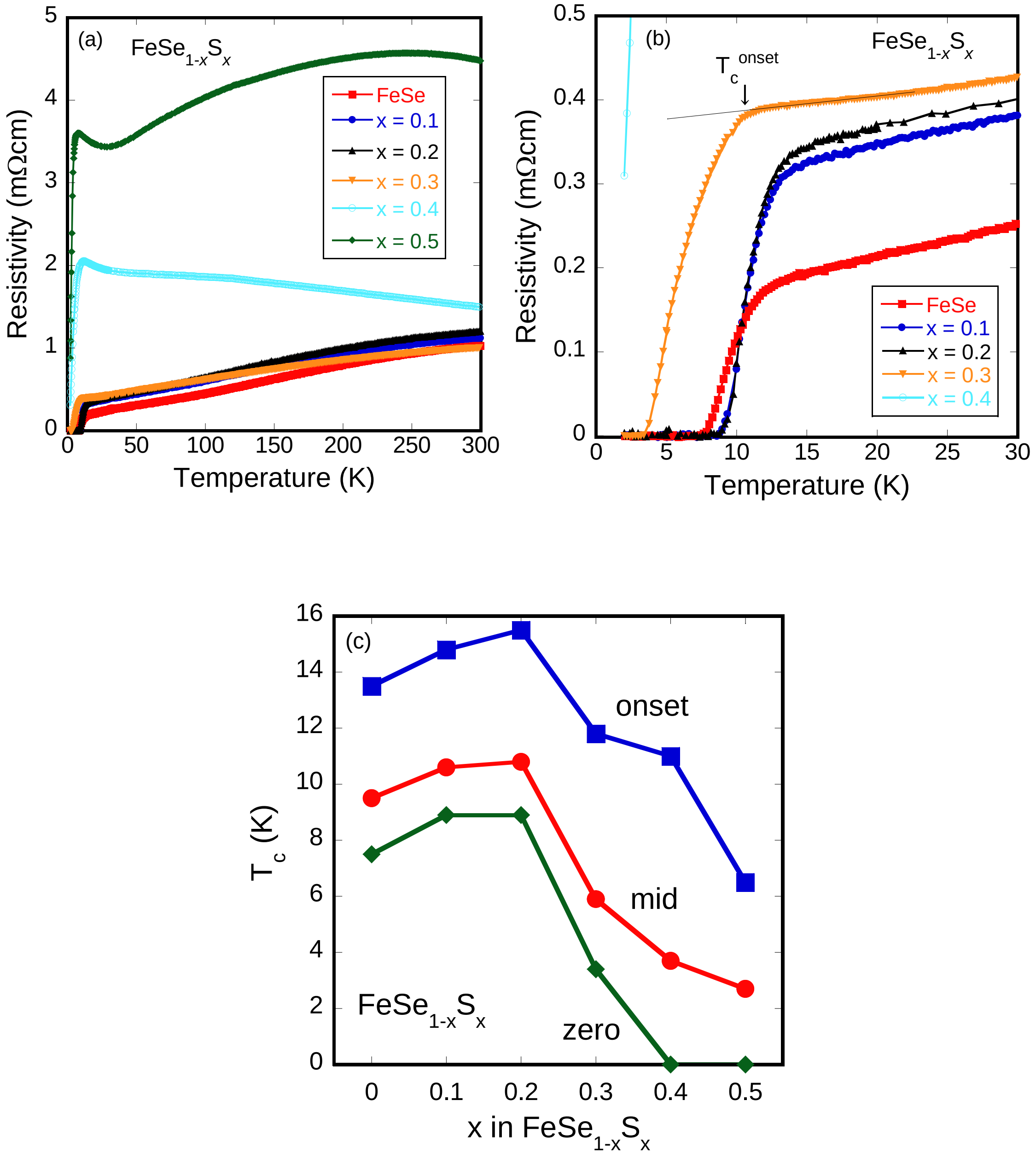

Fig. 4

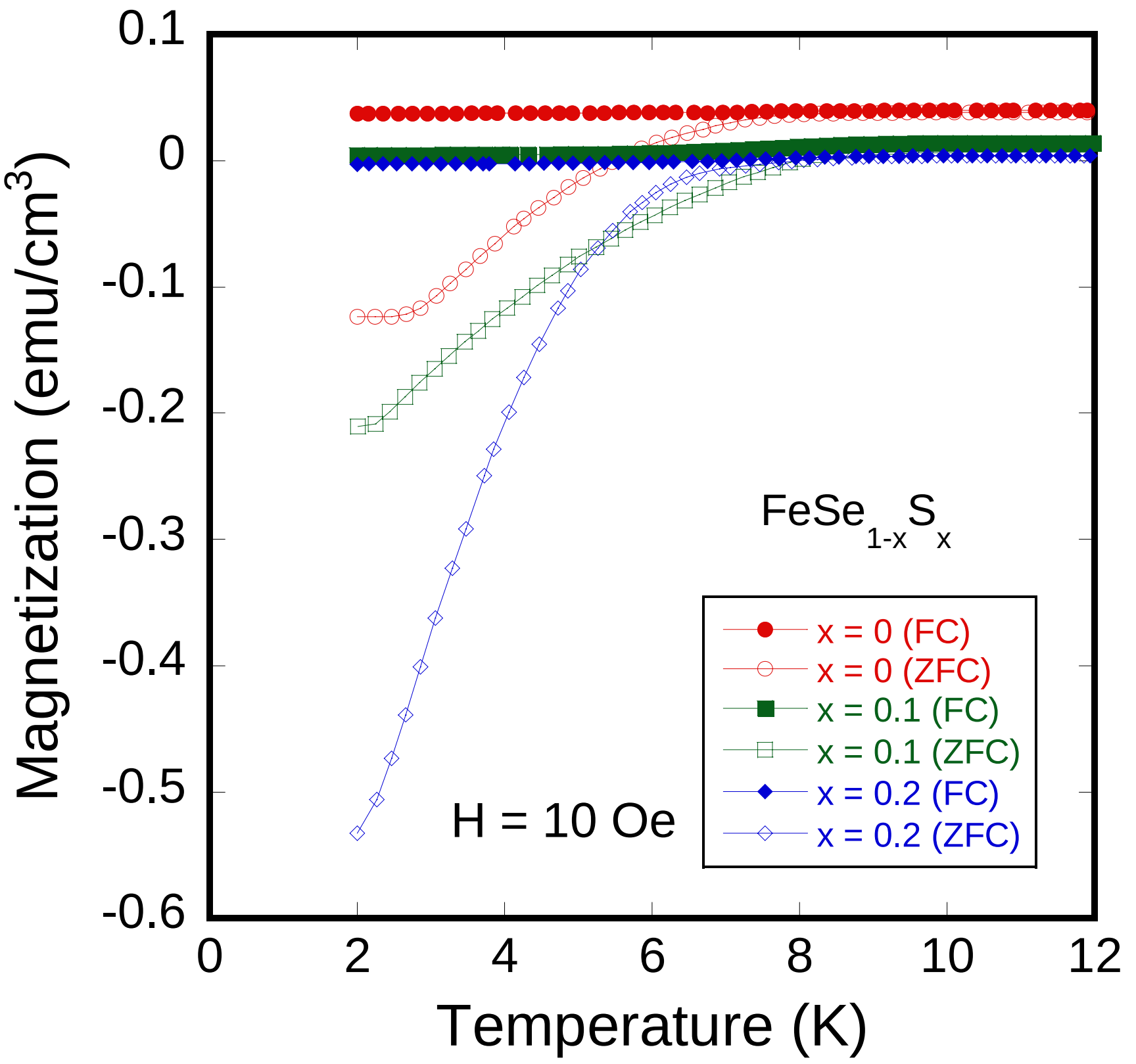

Magnetization (emu/cm3)
Temperature (K)
0.1
0
-0.1
-0.2
-0.3
-0.4
-0.5
-0.6
0
2
4
6
8
10
12
FeSe1-xSx
H = 10 Oe
x = 0 (FC)
x = 0 (ZFC)
x = 0.1 (FC)
x = 0.1 (ZFC)
x = 0.2 (FC)
x = 0.2 (ZFC)

Fig. 5

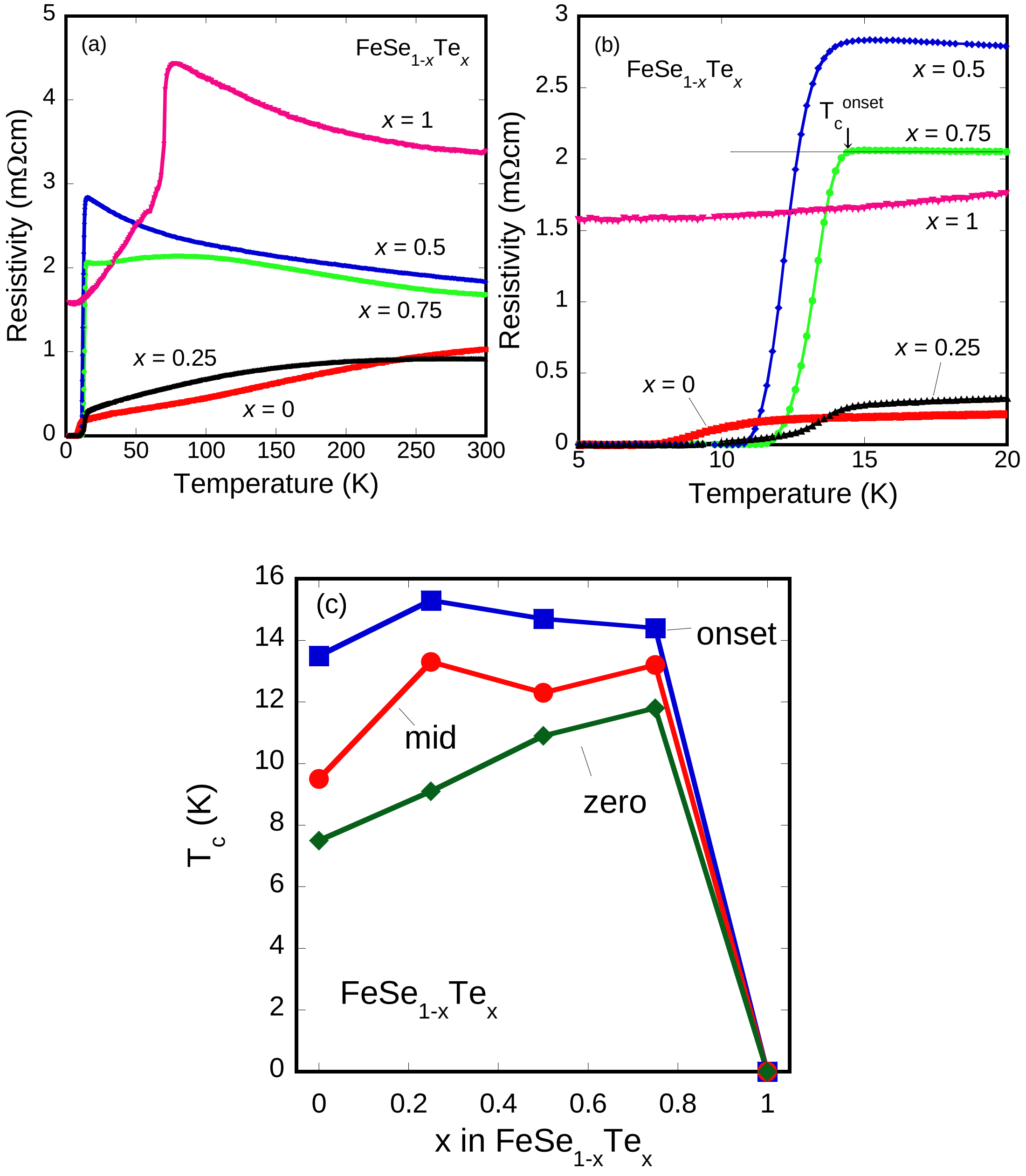

Fig. 6

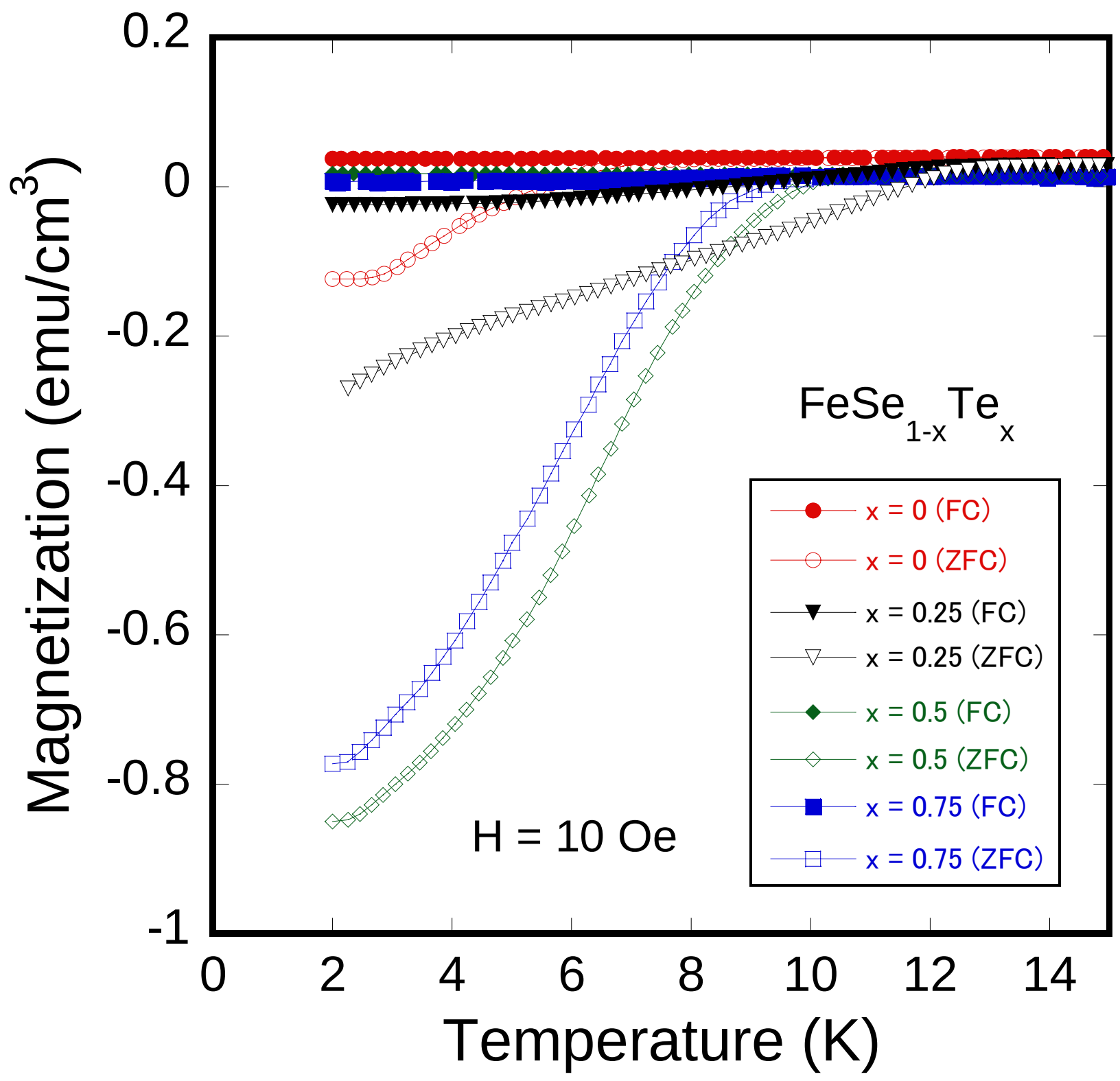

0.2
0
-0.2
-0.4
-0.6
-0.8
-1
Magnetization (emu/cm$^3$)
0
2
4
6
8
10
12
14
Temperature (K)
FeSe$_{1-x}$Te$_x$
x = 0 (FC)
x = 0 (ZFC)
x = 0.25 (FC)
x = 0.25 (ZFC)
x = 0.5 (FC)
x = 0.5 (ZFC)
x = 0.75 (FC)
x = 0.75 (ZFC)
H = 10 Oe

Fig. 7

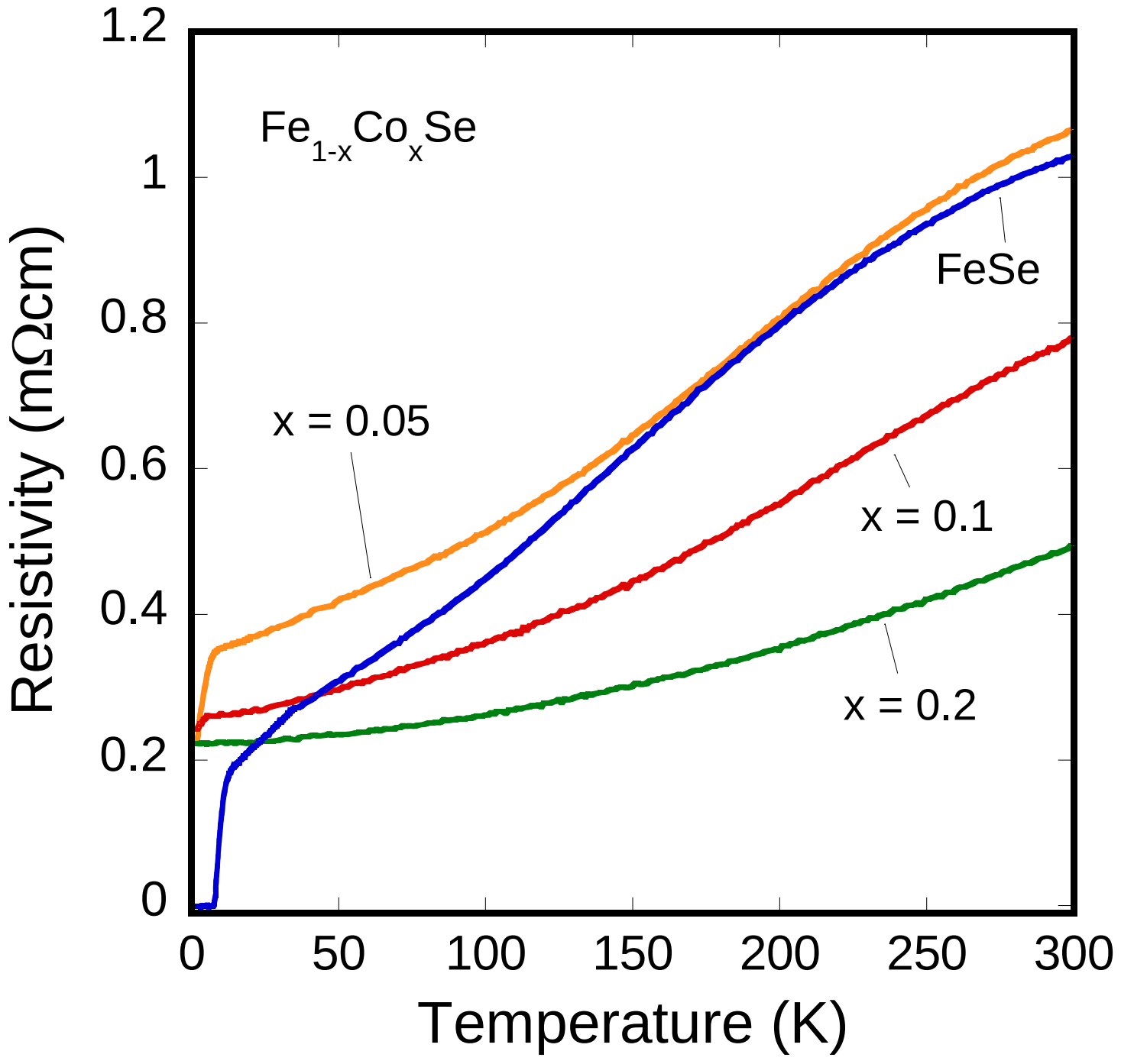

Fig. 8

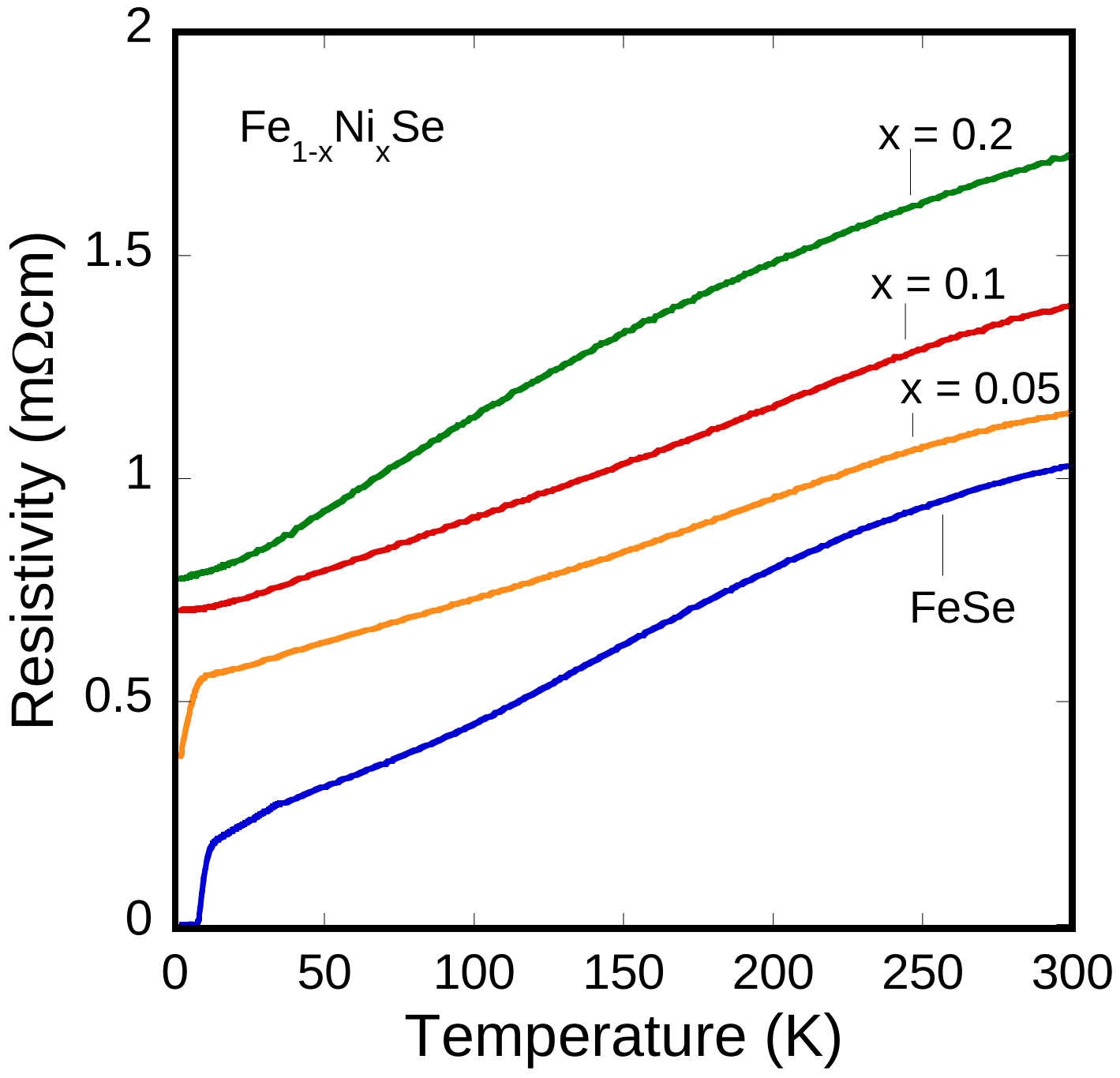

Fig. 9

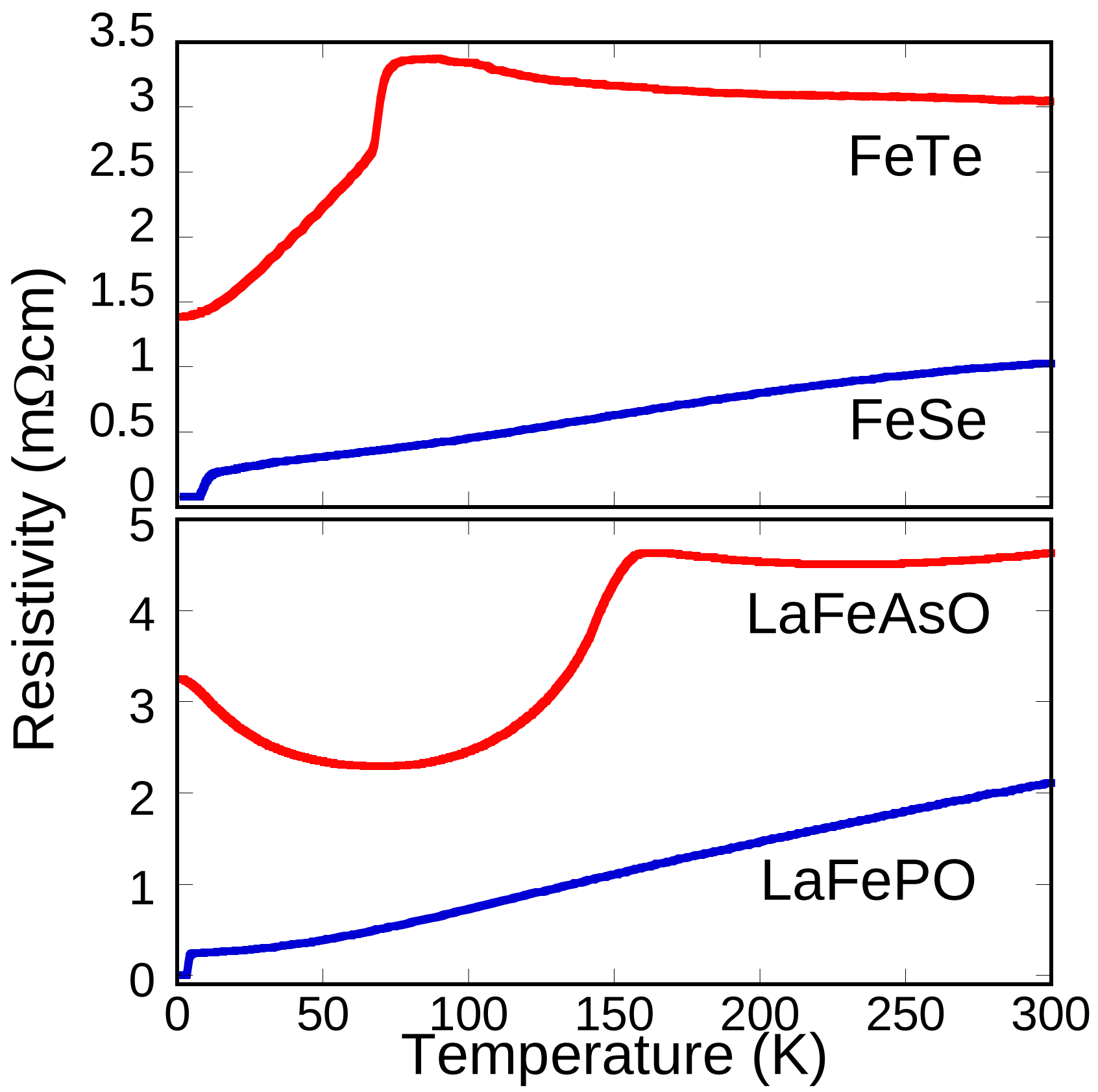